\documentclass{elsart}

\usepackage{natbib}

 \usepackage{graphicx}

\usepackage{amssymb}

\begin{document}

\begin{frontmatter}



\title{Global monitoring of tropospheric water vapor with GPS radio occultation aboard CHAMP}


\author{S. Heise},
\ead{stefan.heise@gfz-potsdam.de}
\author{J. Wickert},
\author{G. Beyerle},
\author{T. Schmidt},
\author{and Ch. Reigber}

\address{GeoForschungsZentrum Potsdam (GFZ), Dep. Geodesy and Remote Sensing, Telegrafenberg, Potsdam, Germany}

\begin{abstract}
The paper deals with application of GPS radio occultation (RO)
measurements aboard CHAMP for the retrieval of tropospheric water
vapor profiles. The GPS RO technique provides a powerful tool for
atmospheric sounding which requires no calibration, is not
affected by clouds, aerosols or precipitation, and provides an
almost uniform global coverage. We briefly overview data
processing and retrieval of vertical refractivity, temperature and
water vapor profiles from GPS RO observations. CHAMP RO data are
available since 2001 with up to 200 high resolution atmospheric
profiles per day. Global validation of CHAMP water vapor profiles
with radiosonde data reveals a bias of about 0.2 g/kg and a
standard deviation of less than 1 g/kg specific humidity in the
lower troposphere. We demonstrate potentials of CHAMP RO
retrievals for monitoring the mean tropospheric water vapor
distribution on a global scale.
\end{abstract}

\begin{keyword}
CHAMP, GPS radio occultation, water vapor, troposphere
\end{keyword}
\end{frontmatter}

\section{Introduction}\label{sec:intro}
The implementation of the Global Positioning System (GPS) enabled
the development of the radio occultation (RO) technique
\citep[e.g.][]{melb94, kurs97} for remote sensing of the Earth's
atmosphere. This technique exploits atmospheric refraction of GPS
signals observed aboard Low Earth Orbiting (LEO) satellites. Basic
observable is the atmospheric excess phase which is used for the
retrieval of meteorological quantities. The potentials of GPS RO
measurements for providing vertical atmospheric profiles of
refractivity, temperature and water vapor have been demonstrated
for the first time by the GPS/MET experiment
\citep[e.g.][]{ware96, kurs97}. The CHAMP (Challenging
Minisatellite Payload) GPS RO experiment was successfully started
on Feb. 11, 2001 \citep{wick01} and is activated continuously
since mid of 2001. Considering current and planned LEO satellite
missions (e.g. GRACE or COSMIC), GPS RO data will provide a
valuable data base for climatological investigations and
improvement of global weather forecasts in the future. One
challenge in processing GPS radio occultation measurements is the
data analysis in the lower troposphere. The refractivity retrieved
from GPS RO data shows a negative bias in relation to
meteorological data \citep{ao03, bey04}, which leads to a
corresponding bias in the retrieved specific humidity. Reasons for
this bias are GPS receiver tracking errors, uncorrected multipath
in signal propagation and critical refraction. The application of
advanced retrieval techniques, as the Full Spectrum Inversion
(FSI) method \citep{jen03}, reduces the refractivity bias
significantly \citep{wick04}. The FSI is implemented to the
current version (005) of the operational data analysis software at
GFZ Potsdam, which is available since February 2004.

\section{Retrieval technique}\label{sec:retrieval}
The retrieval of atmospheric profiles from GPS occultation
measurements has been described in detail by a number of authors
\citep[e.g.][]{melb94, kurs97, wick04}. Briefly, the GPS
measurements recorded by a receiver onboard a LEO (50 Hz sampling
rate in case of CHAMP) are used together with high precision orbit
information (LEO and occulting GPS satellite) to derive the
atmospheric excess phase with millimetric accuracy. These data are
transformed into profiles of the ray path bending angle
$\alpha${\it (a)}, where {\it a} denotes the so-called impact
parameter. Assuming spherical symmetry the Abel transform
\citep[e.g.][]{fjeld71} is applied to invert $\alpha${\it (a)}
into refractivity profiles $ N(r) $, where $ r $ denotes the ray
path tangential altitude.

\begin{equation}
    N(r) = 77.6\frac{p}{T}+3.73\cdot10^5\frac{p_w}{T^2} \label{eq:ref}
\end{equation}

The atmospheric refractivity $ N(r) $ (Eq. \ref{eq:ref}) is
related to air pressure ($ p $), air temperature ($ T $) and water
vapor pressure ($ p_w $).  Assuming dry air conditions,
refractivity is direct proportional to air density and the
pressure profile can be derived by downward integration of the
refractivity profile assuming hydrostatic equilibrium. The
temperature profile is calculated consecutively by Eq.
\ref{eq:ref}. The validation of CHAMP dry temperature profiles
with radiosonde data as well as ECMWF analyses shows a temperature
bias less than 0.5 K (RMS deviation 1--2 K) between 250--20 hPa
\citep{wick04}.
\begin{figure}
\centering
\includegraphics[height=4.4cm]{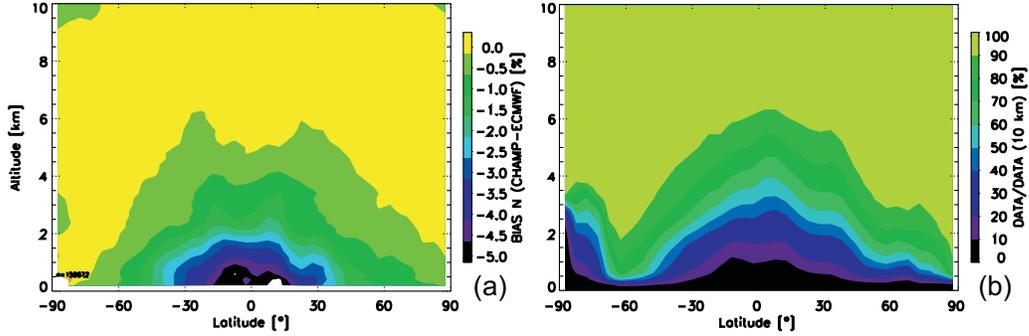}
\caption{(a): Statistical comparison (zonal mean) of
  CHAMP refractivity (product version 005,
  159,672 profiles, May 2001 to August 2004) with ECMWF. (b): Number of profiles in
  relation to profile availability at 10 km altitude corresponding to left panel.}
\label{fig:1}       
\vspace*{0mm}
\end{figure}

However, the dry air assumption is not valid over wide areas of
the mid and lower troposphere, leading to an ambiguity of the
'dry' and 'wet' refractivity term (Eq. \ref{eq:ref}). To deal with
this problem, additional meteorological information is necessary.
There are different retrieval techniques using such information.
\citet {gorb93} describe an iterative algorithm which uses
external temperature information (e.g. from ECMWF analyses) to
separate dry and wet part of the measured refractivity. In result,
pressure and humidity profile are derived from the refractivity
data. Another approach to retrieve both humidity and temperature
profile is the 1Dvar technique \citep{healy00}. This optimal
estimation method requires background information (temperature,
humidity and pressure) as well as error characteristics of the
measurement (refractivity) and the background (e.g. ECMWF). A
further approach, in the following referred to as direct water
vapor pressure (DWVP) retrieval, has been developed at GFZ
Potsdam. Here background (ECMWF) temperature and pressure
information are used to calculate water vapor pressure ($ p_w $)
directly form refractivity data applying Eq. \ref{eq:ref}. The
difference between the so derived $ p_w $ and background humidity
information is used to adapt the background pressure for a
recalculation of $ p_w $. The pressure values converge very
quickly and the procedure is stopped after the second iteration
step. The known negative refractivity bias in the lower
troposphere (Fig. \ref{fig:1}(a)) states a general problem for the
humidity derivation. To avoid DWVP retrieval outliers especially
in that region, deviations between background and retrieval
humidity are restricted to the double ECMWF humidity error of the
current profile during the iteration.

At GFZ Potsdam both, 1Dvar and DWVP algorithms are implemented for
tropospheric water vapor retrieval. In the following we present
results from both techniques and compare these with radiosonde
measurements.

\section{Results and validation}\label{sec:res}
\begin{figure}
\centering
\includegraphics[height=6.4cm]{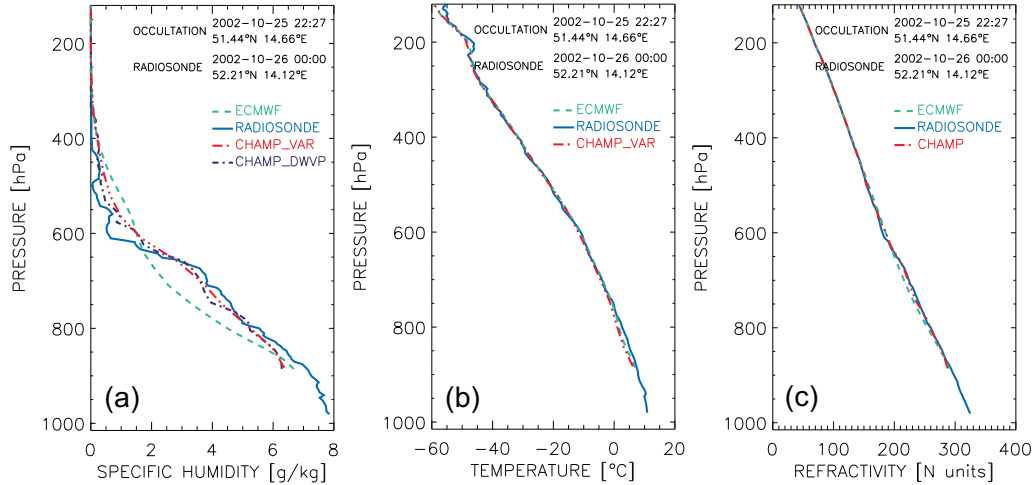}
\caption{Comparison of vertical specific humidity
 (a), temperature (b) and  refractivity(c) profiles derived from CHAMP
  DWVP and 1Dvar retrieval with radiosonde Lindenberg and ECMWF data. Example
  for occultation 226, October 25, 2002, 22:27 UTC, 51.44$^\circ$N, 14.66$^\circ$E.}
\label{fig:2}       
\vspace*{0mm}
\end{figure}

To give an impression on the retrieval results, Fig.
\ref{fig:2}(a) shows an example of CHAMP 1Dvar and DWVP specific
humidity in comparison to radiosonde and ECMWF data. Both CHAMP
retrievals come to quite similar results revealing significant
improvement of the background (ECMWF) specific humidity in
comparison to radiosonde data. Nevertheless, 1Dvar shows a
slightly smoother result than DWVP. This has been observed in
several cases and can be considered as a general difference
between both retrievals. Obviously, DWVP is more sensitive to
vertical structures in the input refractivity than 1Dvar. Fig.
\ref{fig:2}(b) reveals good agreement between radiosonde and ECMWF
temperature profiles. The 1Dvar temperature shows only small
deviations from the background. Finally, radiosonde, ECMWF and
CHAMP refractivity profiles are given in Fig. \ref{fig:2}(c).
Especially between 600 and 800 hPa the CHAMP refractivity shows
better agreement to the radiosonde observation than ECMWF. This
obviously corresponds to improvements of the CHAMP humidity
retrieval above the background humidity in this altitude range.

\begin{figure}
\centering
\includegraphics[height=5.cm]{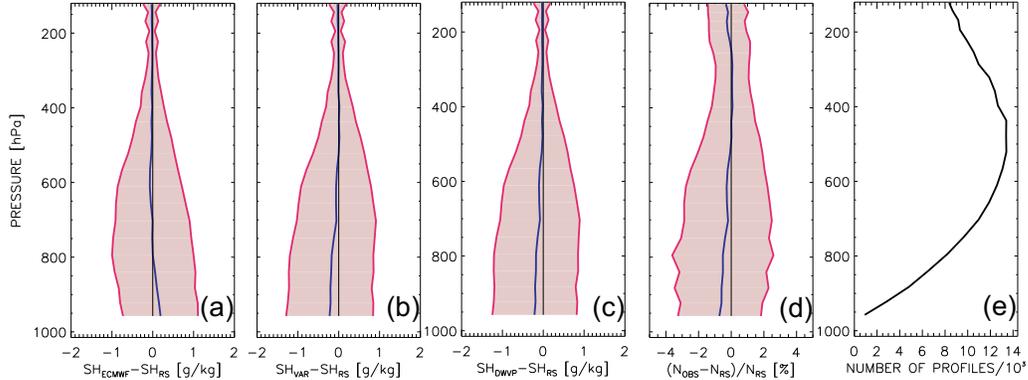}
\caption{Statistical comparison (years 2002-2003) of vertical
specific humidity profiles from global radiosonde stations with:
(a) ECMWF, (b) 1Dvar, (c) DWVP. Corresponding comparison of CHAMP
and radiosonde refractivity is shown in (d). Blue lines represent
bias, red lines standard deviation. Number of compared data points
(coincidence radius: 300 km, 3 hours) is shown in (e).}
\label{fig:3}       
\vspace*{0mm}
\end{figure}

The 1Dvar and DWVP retrieval results have been validated with
radiosonde data. Fig. \ref{fig:3} (a)-(c) show the statistical
comparison (bias and standard deviation) of vertical specific
humidity profiles from ECMWF, 1Dvar and DWVP with coinciding
radiosonde profiles on a global scale. For the years 2002 and 2003
about 13,400 coincidences have been found (see Fig. \ref{fig:3}
(e), coincidence radius: 300 km spatial and 3 hours temporal).
Radiosonde data were quality checked by comparison with ECMWF and
have been ignored in case of more than 10\% refractivity
deviation. As can be seen from Fig. \ref{fig:3}(b) and (c), the
statistical comparison of 1Dvar and DWVP with radiosonde data
comes to quite similar results. The negative CHAMP refractivity
bias (Fig. \ref{fig:3}(d)) leads to a specific humidity bias of of
about -0.2 g/kg in the lower troposphere. Nevertheless, the
standard deviation is similar to the result from radiosonde
comparison with ECMWF (Fig. \ref{fig:3}(a)). By comparing Fig.
\ref{fig:3}(a)-(d) it has to be noticed that between 600 and 500
hPa DWVP shows a slightly stronger relation to the refractivity
data than 1Dvar which seems to be more influenced by the
background (ECMWF) data in this region.

\section{Global application}\label{sec:appl}
\begin{figure}
\centering
\includegraphics[height=2.9cm]{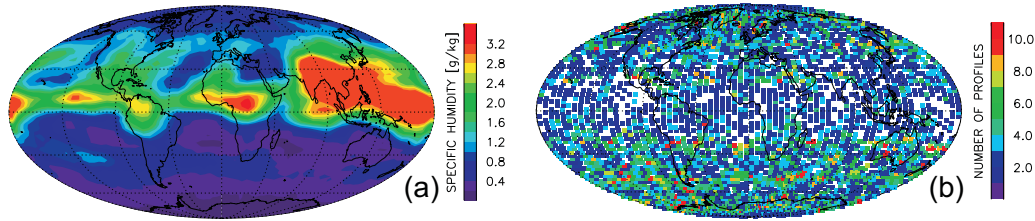}
\caption{(a): CHAMP DWVP mean global water vapor
  distribution at 500 hPa for northern summer conditions
  (June--August 2002). (b): Data coverage according to (a) (12575 data points).}
\label{fig:4}       
\vspace*{0mm}
\end{figure}

The CHAMP humidity profiles may be used for investigation of mean
seasonal or medium term water vapor distribution on a global
scale. Due to rather low data exploitation in the lower
troposphere (see Fig. \ref{fig:1}(b) and Fig. \ref{fig:3}(e)) and
accuracy restrictions at low humidity levels in the upper
troposphere, the mid troposphere region is most appropriate for
such investigations. Fig. \ref{fig:4}(a) shows the mean global
water vapor distribution at 500 hPa pressure level derived from
DWVP results for the northern summer season of 2002 according to a
grid of 2.5$^\circ$ resolution in latitude and 5.0$^\circ$ in
longitude respectively. The corresponding data coverage (profiles
per pixel) is shown in \ref{fig:4}(b). It has to be mentioned that
current and planned LEO RO missions like GRACE and COSMIC will
provide a significantly extended data base which will allow for
global coverage within much shorter time scales than the CHAMP
mission alone. Furthermore, RO data will be of growing interest
for climatological investigations in the future if first medium
and long term RO data sets become available.
\section{Conclusions}\label{sec:con}
The 1Dvar and DWVP techniques state valuable tools for the
humidity retrieval from occultation refractivity measurements in
the mid and lower troposphere. Statistically, both methods come to
comparable results. Even if a unique separation of dry and wet
refractivity components is not possible, the derived humidity
information may be improved above the background (e.g. ECMWF)
provided that temperature background and refractivity measurement
are of sufficient quality. Advanced retrieval techniques like the
FSI method significantly reduced the negative refractivity bias in
the lower troposphere. Nevertheless, the refractivity retrieval
needs further improvement. Potentials of CHAMP RO data for global
water vapor monitoring have been demonstrated. The CHAMP RO
experiment generates the first long-term RO data set. Following
satellite missions like GRACE and COSMIC will significantly extend
the RO database improving the capabilities for global water vapor
monitoring and climatological investigations.

{\bf Acknowledgements}

Many thanks to Sean Healy for helpful comments and instructions
concerning the implementation of 1Dvar. We thank Katrin
Sch\"{o}llhammer and the Institute for Meteorology at the Free
University Berlin for delivering radiosonde data and the ECMWF for
supplying global weather analyses.



\end{document}